%% file: main.tex
\errorstopmode
\input amssym.def
\input amssym.tex
\input format
\input epsf
\epsfclipon
\input macros

\input title

\input sect1

\input sect2

\input sect3

\input sect4

\input appa

\input appb

\input biblio

\bye

%% file: format

\magnification=\magstephalf
\hsize=14.0 true cm
\vsize=19 true cm
\hoffset=1.0 true cm
\voffset=2.0 true cm

\abovedisplayskip=12pt plus 3pt minus 3pt
\belowdisplayskip=12pt plus 3pt minus 3pt
\parindent=1.0em


\font\sixrm=cmr6
\font\eightrm=cmr8
\font\ninerm=cmr9

\font\sixi=cmmi6
\font\eighti=cmmi8
\font\ninei=cmmi9

\font\sixsy=cmsy6
\font\eightsy=cmsy8
\font\ninesy=cmsy9

\font\sixbf=cmbx6
\font\eightbf=cmbx8
\font\ninebf=cmbx9

\font\eightit=cmti8
\font\nineit=cmti9

\font\eightsl=cmsl8
\font\ninesl=cmsl9

\font\sixss=cmss8 at 8 true pt
\font\sevenss=cmss9 at 9 true pt
\font\eightss=cmss8
\font\niness=cmss9
\font\tenss=cmss10

\font\sixmib=cmmib6
\font\sevenmib=cmmib7
\font\eightmib=cmmib8
\font\ninemib=cmmib9
\font\tenmib=cmmib10

 at 12 true pt
 at 12 true pt
\font\bigrm=cmr10 at 12 true pt
 at 12 true pt
 at 12 true pt

 at 16 true pt
 at 16 true pt
\font\Bigrm=cmr12 at 16 true pt
 at 16 true pt
 at 16 true pt

\catcode`@=11
\newfam\ssfam
\newfam\mibfam

\def\tenpoint{\def\rm{\fam0\tenrm}%
    \textfont0=\tenrm \scriptfont0=\sevenrm \scriptscriptfont0=\fiverm
    \textfont1=\teni  \scriptfont1=\seveni  \scriptscriptfont1=\fivei
    \textfont2=\tensy \scriptfont2=\sevensy \scriptscriptfont2=\fivesy
    \textfont3=\tenex \scriptfont3=\tenex   \scriptscriptfont3=\tenex
    \textfont\itfam=\tenit                  \def\it{\fam\itfam\tenit}%
    \textfont\slfam=\tensl                  \def\sl{\fam\slfam\tensl}%
    \textfont\bffam=\tenbf \scriptfont\bffam=\sevenbf
                           \scriptscriptfont\bffam=\fivebf
                           \def\bf{\fam\bffam\tenbf}%
    \textfont\ssfam=\tenss \scriptfont\ssfam=\sevenss
                           \scriptscriptfont\ssfam=\sevenss
                           \def\ss{\fam\ssfam\tenss}%
    \textfont\mibfam=\tenmib \scriptfont\mibfam=\sevenmib
                             \scriptscriptfont\mibfam=\sevenmib
                             \def\mib{\fam\mibfam\tenmib}%
    \normalbaselineskip=13pt
    \setbox\strutbox=\hbox{\vrule height8.5pt depth3.5pt width0pt}%
    \let\big=\tenbig
    \normalbaselines\rm}

\def\ninepoint{\def\rm{\fam0\ninerm}%
    \textfont0=\ninerm      \scriptfont0=\sixrm
                            \scriptscriptfont0=\fiverm
    \textfont1=\ninei       \scriptfont1=\sixi
                            \scriptscriptfont1=\fivei
    \textfont2=\ninesy      \scriptfont2=\sixsy
                            \scriptscriptfont2=\fivesy
    \textfont3=\tenex       \scriptfont3=\tenex
                            \scriptscriptfont3=\tenex
    \textfont\itfam=\nineit \def\it{\fam\itfam\nineit}%
    \textfont\slfam=\ninesl \def\sl{\fam\slfam\ninesl}%
    \textfont\bffam=\ninebf \scriptfont\bffam=\sixbf
                            \scriptscriptfont\bffam=\fivebf
                            \def\bf{\fam\bffam\ninebf}%
    \textfont\ssfam=\niness \scriptfont\ssfam=\sixss
                            \scriptscriptfont\ssfam=\sixss
                            \def\ss{\fam\ssfam\niness}%
    \textfont\mibfam=\ninemib \scriptfont\mibfam=\sixmib
                            \scriptscriptfont\mibfam=\sixmib
                            \def\mib{\fam\mibfam\ninemib}%
    \normalbaselineskip=12pt
    \setbox\strutbox=\hbox{\vrule height8.0pt depth3.0pt width0pt}%
    \let\big=\ninebig
    \normalbaselines\rm}

\def\eightpoint{\def\rm{\fam0\eightrm}%
    \textfont0=\eightrm      \scriptfont0=\sixrm
                             \scriptscriptfont0=\fiverm
    \textfont1=\eighti       \scriptfont1=\sixi
                             \scriptscriptfont1=\fivei
    \textfont2=\eightsy      \scriptfont2=\sixsy
                             \scriptscriptfont2=\fivesy
    \textfont3=\tenex        \scriptfont3=\tenex
                             \scriptscriptfont3=\tenex
    \textfont\itfam=\eightit \def\it{\fam\itfam\eightit}%
    \textfont\slfam=\eightsl \def\sl{\fam\slfam\eightsl}%
    \textfont\bffam=\eightbf \scriptfont\bffam=\sixbf
                             \scriptscriptfont\bffam=\fivebf
                             \def\bf{\fam\bffam\eightbf}%
    \textfont\ssfam=\eightss \scriptfont\ssfam=\sixss
                             \scriptscriptfont\ssfam=\sixss
                             \def\ss{\fam\ssfam\eightss}%
    \textfont\mibfam=\eightmib \scriptfont\mibfam=\sixmib
                             \scriptscriptfont\mibfam=\sixmib
                             \def\mib{\fam\mibfam\eightmib}%
    \normalbaselineskip=10pt
    \setbox\strutbox=\hbox{\vrule height7.0pt depth2.0pt width0pt}%
    \let\big=\eightbig
    \normalbaselines\rm}

\def\tenbig#1{{\hbox{$\left#1\vbox to8.5pt{}\right.\n@space$}}}
\def\ninebig#1{{\hbox{$\textfont0=\tenrm\textfont2=\tensy
                       \left#1\vbox to7.25pt{}\right.\n@space$}}}
\def\eightbig#1{{\hbox{$\textfont0=\ninerm\textfont2=\ninesy
                       \left#1\vbox to6.5pt{}\right.\n@space$}}}

\font\sectionfont=cmbx10
\font\subsectionfont=cmti10

\def\figurecaptionfont{\ninepoint}
\def\tablecaptionfont{\ninepoint}
\def\footnotefont{\eightpoint}


\newcount\equationno
\newcount\bibitemno
\newcount\figureno
\newcount\tableno

\equationno=0
\bibitemno=0
\figureno=0
\tableno=0


\footline={\ifnum\pageno=0{\hfil}\else
{\hss\rm\the\pageno\hss}\fi}


\def\section #1. #2 \par
{\vskip0pt plus .10\vsize\penalty-100 \vskip0pt plus-.10\vsize
\vskip 1.6 true cm plus 0.2 true cm minus 0.2 true cm
\global\def\equationlabel{#1}
\global\equationno=0
\leftline{\sectionfont #1. #2}\par
\immediate\write\terminal{Section #1. #2}
\vskip 0.7 true cm plus 0.1 true cm minus 0.1 true cm
\noindent}


\def\subsection #1 \par
{\vskip0pt plus 0.8 true cm\penalty-50 \vskip0pt plus-0.8 true cm
\vskip2.5ex plus 0.1ex minus 0.1ex
\leftline{\subsectionfont #1}\par
\immediate\write\terminal{Subsection #1}
\vskip1.0ex plus 0.1ex minus 0.1ex
\noindent}


\def\appendix #1. #2 \par
{\vskip0pt plus .10\vsize\penalty-100 \vskip0pt plus-.10\vsize
\vskip 1.6 true cm plus 0.2 true cm minus 0.2 true cm
\global\def\equationlabel{\hbox{\rm#1}}
\global\equationno=0
\leftline{\sectionfont Appendix #1. #2}\par
\immediate\write\terminal{Appendix #1. #2}
\vskip 0.7 true cm plus 0.1 true cm minus 0.1 true cm
\noindent}



\def\equation#1{$$\displaylines{\qquad #1}$$}
\def\enum{\global\advance\equationno by 1
\hfill\llap{{\rm(\equationlabel.\the\equationno)}}}

\def\nexteq#1{\cr\noalign{\vskip#1}\qquad}


\def\ifundefined#1{\expandafter\ifx\csname#1\endcsname\relax}

\def\ref#1{\ifundefined{#1}?\immediate\write\terminal{unknown reference
on page \the\pageno}\else\csname#1\endcsname\fi}

\newwrite\terminal
\newwrite\bibitemlist

\def\bibitem#1#2\par{\global\advance\bibitemno by 1
\immediate\write\bibitemlist{\string\def
\expandafter\string\csname#1\endcsname
{\the\bibitemno}}
\item{[\the\bibitemno]}#2\par}

\def\beginbibliography{
\vskip0pt plus .15\vsize\penalty-100 \vskip0pt plus-.15\vsize
\vskip 1.2 true cm plus 0.2 true cm minus 0.2 true cm
\leftline{\sectionfont References}\par
\immediate\write\terminal{References}
\immediate\openout\bibitemlist=biblist
\frenchspacing\parindent=1.8em
\vskip 0.5 true cm plus 0.1 true cm minus 0.1 true cm}

\def\endbibliography{
\immediate\closeout\bibitemlist
\nonfrenchspacing\parindent=1.0em}


\def\figurecaption#1{\global\advance\figureno by 1
\narrower\figurecaptionfont
Fig.~\the\figureno. #1}

\def\tablecaption#1{\global\advance\tableno by 1
\vbox to 0.25 true cm { }
\centerline{\tablecaptionfont%
Table~\the\tableno. #1}
\vskip-0.4 true cm}

\def\thicktablerule{\hrule height0.8pt}
\def\thintablerule{\hrule height0.4pt}

\tenpoint

\immediate\openin\bibitemlist=biblist
\ifeof\bibitemlist\immediate\closein\bibitemlist
\else\immediate\closein\bibitemlist
\input biblist \fi


\def\thismonth{\ifcase\month\or
January\or February\or March\or April\or May\or June\or
July\or August\or September\or October\or November\or December\fi}

%% file: macros


\def\rmd{{\rm d}}

\def\rmO{{\rm O}}



\def\proof{\noindent{\sl Proof:}\kern0.6em}

\def\frac#1#2{\hbox{$#1\over#2$}}
\def\dual{\mathstrut^*\kern-0.1em}

\def\lvec#1{\setbox0=\hbox{$#1$}
    \setbox1=\hbox{$\scriptstyle\leftarrow$}
    #1\kern-\wd0\smash{
    \raise\ht0\hbox{$\raise1pt\hbox{$\scriptstyle\leftarrow$}$}}
    \kern-\wd1\kern\wd0}
\def\rvec#1{\setbox0=\hbox{$#1$}
    \setbox1=\hbox{$\scriptstyle\rightarrow$}
    #1\kern-\wd0\smash{
    \raise\ht0\hbox{$\raise1pt\hbox{$\scriptstyle\rightarrow$}$}}
    \kern-\wd1\kern\wd0}
\def\slash#1{\setbox0=\hbox{$#1$}\setbox1=\hbox{$\kern1pt/$}
    #1\kern-\wd0\kern1pt/\kern-\wd1\kern\wd0}


\def\nabstar#1{{\nabla\kern0.5pt\smash{\raise 4.5pt\hbox{$\ast$}}
               \kern-5.5pt_{#1}}}

\def\drvstar#1{{\partial\kern0.5pt\smash{\raise 4.5pt\hbox{$\ast$}}
               \kern-6.0pt_{#1}}}

\def\ldrvstar#1{{\lvec{\,\partial}\kern-0.5pt\smash{\raise 4.5pt\hbox{$\ast$}}
               \kern-5.0pt_{#1}}}


\def\MeV{{\rm MeV}}

\def\fm{{\rm fm}}
\def\MSbar{\overline{\rm MS\kern-0.5pt}\kern0.5pt}




\def\dirac#1{\gamma_{#1}}
\def\diracstar#1#2{
    \setbox0=\hbox{$\gamma$}\setbox1=\hbox{$\gamma_{#1}$}
    \gamma_{#1}\kern-\wd1\kern\wd0
    \smash{\raise4.5pt\hbox{$\scriptstyle#2$}}}


\def\SUthree{{\rm SU(3)}}

\def\Tr{{\rm Tr}}


\def\Dm{D_m}
\def\Dmdag{{D_m}^{\kern-2pt\dagger}}
\def\mc{m_{\rm c}}
\def\mq{m_{\rm q}}
\def\ZA{Z_{A}}
\def\ZP{Z_{P}}


\def\csw{c_{\rm sw}}
\def\cA{c_{A}}
\def\bA{b_{A}}
\def\bP{b_{P}}

\def\bPP{b_{P\kern-1pt P}}
\def\bPS{b_{P\kern-1pt S}}

\def\bm{b_m}
\def\bmu{b_{\mu}}


\def\RM{{\Bbb R}_M}
\def\PM{{\Bbb P}_M}
\def\eps{\epsilon}

\def\Xop{{\Bbb X}}
\def\Mstar{M_{\ast}}


\def\mr{m_{\hbox{\sixrm R}}}

\def\nur{\nu_{\hbox{\sixrm R}}}
\def\Mr{M_{\hbox{\sixrm R}}}


\def\eps{\epsilon}

%% file: title
%
\rightline{CERN-PH-TH/2010-173}

\vskip1.5cm 
\centerline{\Bigrm
Universality of the topological susceptibility
}
\vskip0.3cm
\centerline{\Bigrm
in the SU(3) gauge theory}
\vskip 0.6 true cm
\centerline{\bigrm Martin L\"uscher and 
Filippo Palombi}
\vskip1.5ex
\centerline{{\it CERN, Physics Department, 1211 Geneva 23, Switzerland}}
\vskip 0.8 true cm
\thintablerule
\vskip 2.0ex
\ninepoint
\leftline{\bf Abstract}
\vskip 1.0ex\noindent
The definition and computation of the 
topological susceptibility in non-abelian gauge theories
is complicated by the presence of non-integrable short-distance 
singularities. 
Recently, alternative representations of the susceptibility
were discovered,
which are singularity-free
and do not require renormalization.
Such an expression is here studied quantitatively, 
using the lattice formulation of the SU(3) gauge theory
and numerical simulations.
The results confirm the expected scaling
of the susceptibility 
with respect to the lattice spacing and
they also agree, within errors,
with computations of the susceptibility based on the use
of a chiral lattice Dirac operator.
\vskip 2.0ex
\thintablerule

\tenpoint

\vskip-0.3cm

%% file: sect1
\section 1. Introduction

In QCD and other non-abelian gauge theories, the discussion of the
effects of the topological properties of the classical field
space tends to be conceptually non-trivial, because the gauge field
integrated over in the functional integral is, with probability $1$, 
nowhere continuous.
The topological susceptibility, for example, is only formally given by
the two-point function of the topological density at 
zero momentum, unless a prescription is supplied of how exactly
the non-integrable short-distance singularity of the two-point 
function is to be treated.

In lattice gauge theory, the problem was reexamined some time ago
[\ref{ChiWI}--\ref{TopSing}]
starting from a formulation of lattice QCD which preserves 
chiral symmetry. An important result of this work was 
that the topological susceptibility can be written as a
ratio of expectation values of other observables which
remains well-defined in the continuum limit.
A particular choice of regularization is then not required,
i.e.~the new formula provides a universal definition of the 
susceptibility. Moreover, this definition is such that
the anomalous chiral Ward identities are fully respected.

The aim in the present paper is to complement these theoretical 
developments by demonstrating the suitability 
of the universal definition for the computation of the 
topological susceptibility in lattice gauge theory.
In this study, the pure $\SUthree$ gauge theory
is considered and a recently proposed version [\ref{TopProj}] 
of the universal formula is used (see sect.~2).
As far as the feasibility of the calculation is concerned,
the results are however expected to be
directly relevant for QCD too.

%% file: sect2
\section 2. Singularity-free expressions for the topological susceptibility

The formula for the susceptibility
obtained in [\ref{TopProj}] 
is not very complicated, but some preparation is
required to be able to write it down.
From the beginning, the theory is considered
on a finite hypercubic lattice
with spacing $a$, volume $V$ and periodic boundary conditions.
While some particular choices have to be made along the way,
these details are expected to be
irrelevant in the continuum limit in view of
the fact that the expression is renormalized and free of
short-distance singularities.

\subsection 2.1 Spectral-projector formula

The construction starts by adding a multiplet of valence quarks
with bare mass $m_0$ to the theory. On the lattice, the 
added fields are taken to be of the Wilson type 
[\ref{Wilson}] and the associated
massive Dirac operator $\Dm$ is assumed to include the 
Pauli term required for O($a$) improvement
[\ref{SW},\ref{OaImp}] 
(the relevant improvement and renormalization
constants are collected in
appendix A).

The hermitian operator
$\Dmdag\Dm$ has a complete set of orthonormal eigenmodes with 
non-negative eigenvalues $\alpha$. 
On average there are only few eigenvalues
below some threshold $\alpha_{\rm th}$ proportional to 
the square of the valence-quark mass (see fig.~1).
Above the threshold, the spectrum has an approximately
constant density with a slight downward trend
in the range considered in the figure.
Such a trend is absent
in two-flavour QCD [\ref{TopProj}],
but is qualitatively in line 
with the behaviour of the spectral density at 
next-to-leading order of quenched
chiral perturbation theory
[\ref{OsbornEtAl}].

\input figure1

The topological susceptibility
is now given by [\ref{TopProj}]
\equation{
  \chi_t={\langle\Tr\{\PM\}\rangle\over V}
  {\langle\Tr\{\dirac{5}\PM\}\Tr\{\dirac{5}\PM\}\rangle\over
   \langle\Tr\{\dirac{5}\PM\dirac{5}\PM\}\rangle}+\rmO(a^2),
  \enum
}
where $\PM$ denotes the orthogonal projector to
the subspace spanned by the eigenmodes of $\Dmdag\Dm$ with eigenvalues
$\alpha<M^2$. It is taken for granted in this formula that
$M^2$ is above the effective threshold $\alpha_{\rm th}$
of the spectrum and that
the renormalized valence-quark mass $\mr$ as well as the renormalized
value $\Mr$ of $M$ are held fixed when the lattice spacing
is taken to zero (cf.~appendix A).

\subsection 2.2 Alternative expressions

Equation (2.1) derives from a study of the renormalization and symmetry 
properties of the $n$-point correlation functions
of the scalar and pseudo-scalar densities of the valence quarks.
There exist various representations 
of the topological susceptibility of a similar kind,
all having the same continuum limit.
In particular, 
\equation{
  \chi_t={\langle\Tr\{\RM^4\}\rangle\over V}
  {\langle\Tr\{\dirac{5}\RM^2\}\Tr\{\dirac{5}\RM^2\}\rangle\over
   \langle\Tr\{\dirac{5}\RM^2\dirac{5}\RM^2\}\rangle}
  +\rmO(a^2)
  \enum
}
for any function $\RM$ of $\Dmdag\Dm$ which
is equal to unity in the vicinity of the 
spectral threshold and rapidly decaying above
$M^2$. The shape of the function can otherwise
be chosen arbitrarily and
only affects the size of the O($a^2$) 
corrections.

In the numerical work reported later, $\RM$ is set to the rational 
approximation of the projector $\PM$ previously used in 
[\ref{TopProj}] for the computation of the mode number in 
two-flavour QCD. While this function is not exactly equal
to unity in the vicinity of the spectral threshold, the 
effect of the deviation on the calculated values
of the topological susceptibility is totally negligible 
with respect to the statistical errors. 
For the reader's convenience, the function
is given explicitly in appendix B.

%% file: figure1
\topinsert
\vbox{
\vskip0.0cm
\centerline{\epsfxsize=7.0cm\epsfbox{plots/nu.eps}}
\vskip0.5cm
\figurecaption{%
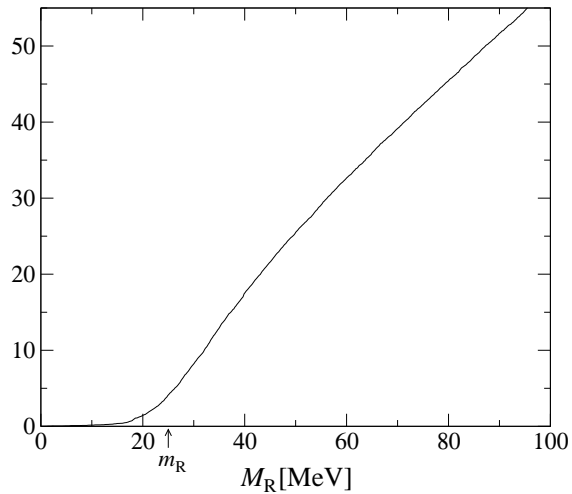
Average number of eigenvalues of $\Dmdag\Dm$ below $M^2$
on a $64\times32^3$ lattice with spacing $a\simeq 0.07$ fm,
plotted as a function of the renormalized value $\Mr$ of $M$.
The renormalized valence-quark mass $\mr$ is about $25$ MeV in this example.
}
}
\endinsert

%% file: sect3
\section 3. Numerical studies

The expression on the right of eq.~(2.2) is a ratio
of well-defined expectation values that
can in principle be computed through numerical simulations. 
In practice, the traces $\Tr\{\ldots\}$ 
can normally not be evaluated exactly,
but as explained in subsect.~3.3,
they can be estimated stochastically
with a moderate computational effort and without
compromising the correctness of the final results.

\input table1

\subsection 3.1 Simulation parameters

The studies reported in this paper
are based on simulations of the lattice theory at 
three values of the inverse
bare gauge coupling $\beta=6/g_0^2$ (see table~1).
A well-known deficit of all currently available simulation 
algorithms for non-abelian gauge theories
(including the link-update algorithms used here)
is the fact that the integrated autocorrelation times of
quantities related to the 
topological charge are rapidly growing when
the lattice spacing decreases
[\ref{DelDebbioTauQ},\ref{StefanConference}].
In order to guarantee the statistical independence 
of the $N_{\rm cnfg}$ gauge-field configurations used
for the ``measurement'' of the topological susceptibility,
the distance in simulation time of subsequent
configurations was required to be at least 10 times larger than 
the relevant autocorrelation times.

Physical units are defined through the
Sommer reference scale $r_0=0.5$ fm [\ref{SommerScale}].
In the range of the gauge coupling covered here, 
the conversion factor $r_0/a$ from lattice to physical units
was accurately determined
by Guagnelli et al.~[\ref{GuagnelliEtAl}].
The spacings of the three lattices thus
decrease from roughly $0.1$ to $0.05$ fm by factors of $1/\sqrt{2}$,
while the lattice sizes in physical units are approximately
constant.

\subsection 3.2 Spectral projector parameters

As already mentioned, the operator $\RM$ is taken to be 
a rational approximation to the projector $\PM$. It thus
depends on the valence-quark mass, the mass $M$ and
the parameters $n$ and $\eps$ that determine the accuracy of
the approximation (cf.~appendix B). A reasonable choice 
of the latter, previously made ref.~[\ref{TopProj}],
is $n=32$ and $\eps=0.01$. In the range of eigenvalues of $\Dmdag\Dm$
below $0.85\times M^2$, the approximation error is then smaller than
$2.2\times10^{-4}$, which is by far small enough to 
guarantee the absence of significant systematic effects 
in eq.~(2.2). Moreover, the contribution of the high modes
is safely suppressed.

The valence-quark mass
and the mass parameter $M$ were adjusted such that their 
renormalized values in the $\MSbar$ scheme at
normalization scale $\mu=2$ GeV 
are about $25$ and $100$ MeV,
respectively.
Using the information collected in appendix A,
the corresponding values of the bare mass parameters,
$\kappa=(8+2am_0)^{-1}$ and $aM$,
can be worked out and are listed in table~1.
On the lattices considered, 
there are then $57-70$ eigenmodes of $\Dmdag\Dm$
with eigenvalues below $M^2$ 
and an average density of roughly
$1$ such mode per $\fm^4$.

As already emphasized, the calculated values of the topological
susceptibility are not expected to strongly depend on all these details
and should in any case always extrapolate to the same 
value in the continuum limit.
The lattice effects are unlikely to be small, however, 
if $aM$ is not much smaller than $1$ or if 
the expectation values on the right of eq.~(2.2) would be 
dominated by the modes up to and slightly above the 
spectral threshold, where the effects are kinematically 
enhanced. Both of these unfavourable situations are
avoided by the above choice of 
the mass parameters.

\subsection 3.3 Random-field representation

In lattice QCD, random field representations
were introduced many years ago [\ref{MichaelPeisa}]
and are now widely used. 
The application of the 
method in the present context 
requires a set
$\eta_1,\ldots,\eta_N$ of $N$ pseudo-fermion fields
to be added to the theory with action
\equation{
  S_{\eta}=\sum_{k=1}^N\left(\eta_k,\eta_k\right),
  \enum
}
where the bracket $(\eta,\zeta)$ denotes the obvious scalar
product of such fields.
For every gauge field configuration,
these fields are generated randomly so that one obtains
a representative ensemble of fields for the complete
theory. In the rest of this section, expectation values
are always taken with respect to both the 
gauge field and pseudo-fermion fields.

The stochastic observables 
\equation{
  {\cal A}={1\over N}\sum_{k=1}^N
  \left(\RM^2\eta_k,\RM^2\eta_k\right),
  \enum
  \nexteq{2ex}
  {\cal B}={1\over N}\sum_{k=1}^N
  \left(\RM\dirac{5}\RM\eta_k,\RM\dirac{5}\RM\eta_k\right),
  \enum
  \nexteq{2ex}
  {\cal C}={1\over N}\sum_{k=1}^N
  \left(\RM\eta_k,\dirac{5}\RM\eta_k\right),
  \enum
}
may now be introduced and a moment of thought then shows that
the expectation values on the right of eq.~(2.2) are given by
\equation{
  \langle\Tr\{\RM^4\}\rangle=
  \langle{\cal A}\rangle,
  \enum
  \nexteq{2ex}
  \langle\Tr\{\dirac{5}\RM^2\}\Tr\{\dirac{5}\RM^2\}\rangle=
  \langle{\cal C}^2\rangle-{\langle{\cal B}\rangle\over N},
  \enum
  \nexteq{2ex}
  \langle\Tr\{\dirac{5}\RM^2\dirac{5}\RM^2\}\rangle=
  \langle{\cal B}\rangle.
  \enum
}
The topological susceptibility can thus be computed by
calculating the expectation values of ${\cal A}$, ${\cal B}$ and 
${\cal C}^2$. For a given gauge-field configuration, the 
evaluation of these observables requires the fields
$\RM\eta_k$, $\RM^2\eta_k$ and $\RM\dirac{5}\RM\eta_k$ to be 
computed, i.e.~the total numerical effort per configuration is
roughly equivalent the one required for $3N$ applications
of the operator $\RM$ to a given pseudo-fermion field.

From this point of view, small values of $N$ are favoured,
but a good choice of $N$ must also take into account
the fact that the variance of the stochastic observables 
decreases with $N$. Some experimenting then shows that 
setting $N=6$ is a reasonable compromise at the specified 
values of the mass parameters. Since $\RM$ is a rational
function of $\Dmdag\Dm$ of degree $[2n+1,2n+1]$, 
the measurement of the stochastic observables requires the 
(twisted-mass) Dirac equation to be solved for altogether $2340$
source fields. 
The computational load thus tends to be heavy, but the problem
is well suited for the application of 
highly efficient solver techniques 
such as local deflation [\ref{LocDefl}]
(see ref.~[\ref{CompStrat}] for a recent review of the subject).
In particular, when these are used, the effort scales
linearly with the lattice size and is nearly independent of 
the values of the mass parameters.

\input table2

\subsection 3.4 Simulation results

The simulation data discussed in the following paragraphs
are summarized in table~2. In all cases,
the statistical errors were estimated
using the jackknife method and
were combined in quadrature with the quoted scale errors
(where appropriate).

\vskip0.5ex
\noindent
(a) {\it Mode number}.
The average number $\nu$ of eigenmodes of $\Dmdag\Dm$ 
with eigenvalues below $M^2$ is an extensive quantity and is
therefore normalized by the lattice volume in table~2.
At the specified bare masses, 
the renormalized masses $\mr$ and $\Mr$ are practically equal to 
$25$ and $100$ MeV, respectively, on all three lattices considered.
In view of the renormalization properties of the mode number
[\ref{TopProj}], the calculated values of $\nu/V$ are thus 
expected to be the same up to O($a^2$) effects.

Within errors,
the figures listed in the second column of table~2 in fact
coincide with one another.
Note that the quoted errors do not take into
account the fact that the mass renormalization factors
and thus the renormalized values of the 
masses are only known up to an error of about $2\%$
(see appendix A). 
Once this error is included in the analysis,
one can still conclude, however, that the simulation results
confirm the expected scaling of the mode number to the continuum limit
at a level of precision of $3\%$ or so.

\vskip0.5ex
\noindent
(b) {\it Topological susceptibility}.
The values of the 
susceptibility calculated along the lines of the present paper 
are listed in the third column of table~2.
Again one observes no statistically significant dependence on 
the lattice spacing. Finite-volume effects are, incidentally, known
to be negligible with respect to the statistical errors on all
three lattices [\ref{ChiGW}]. 

Fits of the data by a constant or a linear function in $a^2$
yield consistent results in the continuum limit. Since the slope 
in $a^2$ turns out to vanish within errors, the (more accurate) number
\equation{
  \chi_t^{1/4}=196.5(5.1)\,\MeV
  \enum
}
obtained by fitting with a constant is quoted here.
This result happens to be practically on top of the
value $194.5(2.4)$ MeV
obtained by Del Debbio, Giusti and Pica 
[\ref{ChiGW}] using a chiral lattice Dirac operator and 
the index theorem [\ref{IndThm}]\kern1pt\footnote{$\dagger$}{\footnotefont%
In ref.~[\ref{ChiGW}], a different convention for the 
conversion from lattice to physical units was used and
the value for the susceptibility quoted there is therefore
slightly different from the one printed here.}.

\vskip0.5ex
\noindent
(c) {\it Charge sectors \& the Wilson flow}.
An understanding of how exactly the topological charge sectors 
emerge in the continuum limit has recently been achieved using
the Wilson flow [\ref{WilsonFlow}]. 
The definition of the topological susceptibility
suggested by the sector division is geometrically appealing
and computationally 
far less demanding than the spectral-projector 
formula (2.2). Presumably the two definitions
agree in the continuum limit, but there is currently
no solid theoretical argument that would show this to be
the case.

The values of the susceptibility computed using the Wilson 
flow are listed in the 
fourth column of table~2 (see ref.~[\ref{WilsonFlow}] 
for the details of the calculation).
While they appear to be systematically lower 
than the ones obtained using the spectral-projector formula,
the differences are statistically insignificant on each lattice.
Moreover, there could be lattice effects of size up to 
the level of the statistical errors.

Since the same ensemble of representative gauge-field configurations
was used in the two cases, the quoted errors are correlated to some extent
(not completely so in view of the use of random fields).
The ratio listed in the last column of table~2 
is therefore obtained with slightly better precision than 
the susceptibilities.
Fits of the ratio
by a constant and linear function in $a^2$ are both possible, 
the values in the continuum limit being $1.048(14)$ and $1.036(31)$, 
respectively. 
The spectral-projector
and the Wilson-flow definition of the susceptibility
thus coincide to a precision of a few percent.
While there is some tension in the data,
there is no clear evidence for the definitions to be
different at this level of accuracy.

%% file: table1
\topinsert
\newdimen\digitwidth
\setbox0=\hbox{\rm 0}
\digitwidth=\wd0
\catcode`@=\active
\def@{\kern\digitwidth}
\tablecaption{Lattice parameters, statistics and bare mass parameters} 
\vskip-1.2ex
$$\vbox{\settabs\+&%
                  xxxxxxxxxxx&xx&
                  xxxxxx&xx&
                  xxxxxxxxxxx&xx&
                  xxxxxxx&xx&
                  xxxxxxxxxx&xx&
                  xxxxxxxxx&x\cr
\thicktablerule
\vskip1.0ex
                \+& \hfill Lattice\hfill
                 && \hfill $\beta$\hfill
                 && \hfill $a$\kern2pt[fm]\hfill
                 && \hfill $N_{\rm cnfg}$\hfill
                 && \hfill $\kappa$\hfill
                 && \hfill $aM$\hfill
                 &\cr
\vskip1.0ex
\thintablerule
\vskip1.2ex
  \+& \hfill $48\times24^3$ \hfill
  &&  \hfill $5.96$ \hfill
  &&  \hfill $0.0999(4)$ \hfill
  &&  \hfill $100$ \hfill
  &&  \hfill $0.134433$ \hfill
  &&  \hfill $0.03188$ \hfill
  &\cr
\vskip0.3ex
  \+& \hfill $64\times32^3$ \hfill
  &&  \hfill $6.17$ \hfill
  &&  \hfill $0.0710(3)$ \hfill
  &&  \hfill $100$ \hfill
  &&  \hfill $0.135540$ \hfill
  &&  \hfill $0.02232$ \hfill
  &\cr
\vskip0.3ex
  \+& \hfill $96\times48^3$ \hfill
  &&  \hfill $6.42$ \hfill
  &&  \hfill $0.0498(3)$ \hfill
  &&  \hfill $100$ \hfill
  &&  \hfill $0.135561$ \hfill
  &&  \hfill $0.01566$ \hfill
  &\cr
\vskip1.2ex
\thicktablerule
}
$$
\endinsert

%% file: table2
\topinsert
\newdimen\digitwidth
\setbox0=\hbox{\rm 0}
\digitwidth=\wd0
\catcode`@=\active
\def@{\kern\digitwidth}
\tablecaption{Mode number and topological susceptibility} 
\vskip-1.2ex
$$\vbox{\settabs\+&%
                  xxxxxx&xx&
                  xxxxxxxxxxx&xx&
                  xxxxxxxxxxxxxx&xx&
                  xxxxxxxxxxxxxx&xx&
                  xxxxxxxxxxxxxx&x\cr
\thicktablerule
\vskip1.0ex
                \+& \hfill $\beta$\hfill
                 && \hfill $\nu/V\,[\fm^{-4}]$\hfill
                 && \hfill 
                 $\chi_t^{1/4}\,[\MeV]$\hfill
                 && \hfill 
                 $(\chi_t^{1/4})_{\hbox{\sixrm WF}}\,[\MeV]$\hfill
                 && \hfill $\chi_t^{1/4}/
                            (\chi_t^{1/4})_{\hbox{\sixrm WF}}$\hfill
                 &\cr
\vskip1.0ex
\thintablerule
\vskip1.2ex
  \+& \hfill $5.96$\hfill
  &&  \hfill $1.053(18)$\hfill
  &&  \hfill $197.3(7.7)$ \hfill
  &&  \hfill $187.7(6.0)$ \hfill
  &&  \hfill $1.051(25)$ \hfill
  &\cr
\vskip0.3ex
  \+& \hfill $6.17$\hfill
  &&  \hfill $1.075(22)$\hfill
  &&  \hfill $203.8(9.1)$ \hfill
  &&  \hfill $192.8(7.0)$ \hfill
  &&  \hfill $1.057(22)$ \hfill
  &\cr
\vskip0.3ex
  \+& \hfill $6.42$\hfill
  &&  \hfill $1.060(24)$\hfill
  &&  \hfill $186.6(9.9)$ \hfill
  &&  \hfill $181.0(7.3)$ \hfill
  &&  \hfill $1.031(26)$ \hfill
  &\cr
\vskip1.2ex
\thicktablerule
}
$$
\endinsert

%% file: sect4
\section 4. Conclusions

The numerical studies reported in this paper confirm the 
universality
of the spectral-projector formula (2.2)
for the topological susceptibility. 
In particular,
no statistically significant lattice-spacing effects were observed
and the calculated values agree
with the result obtained by
Del Debbio, Giusti and Pica [\ref{ChiGW}],
where a chiral lattice Dirac operator was used.

The numerical effort required
for the stochastic evaluation of the spectral-projector
formula increases proportionally to the 
number $V/a^4$ of lattice points, but is 
a flat function of all other parameters.
On large lattices, computations of the susceptibility
along these lines thus tend to be more feasible
than those based on a chiral lattice Dirac operator
(which scale roughly like $V^2$). Even less
computer time is required 
if the susceptibility is defined via the Wilson flow,
but a formal proof for this definition to be in the same 
universality class as the spectral-projector formula 
is still missing.

With respect to the pure gauge theory,
the application of the spectral-projector formula 
in QCD is not expected to run into additional 
difficulties.
Accurate calculations of the topological susceptibility 
however require representative ensembles 
of, say, a few hundred 
statistically independent gauge-field configurations
to be generated.
This part of the calculation usually consumes most of 
the computer time and may rapidly become prohibitively expensive 
at small lattice spacings [\ref{StefanConference}].
At present, computations of the susceptibility on lattices
similar to the ones considered here
are therefore not easily extended to QCD with
light sea quarks.

\vskip0.5ex
We wish to thank Leonardo Giusti for helpful discussions
on various issues related to this work.
All numerical calculations were performed on a
dedicated PC cluster at CERN. We are grateful
to the CERN management for providing the required funds
and to the CERN IT Department for technical support.
F.~P.~acknowledges financial support by an EIF Marie Curie fellowship
of the European Community's Seventh Framework Programme under contract
number PIEF-GA-2009-235904.

%% file: appa
\appendix A. O($\mib a$) improvement and renormalization

\vskip-2.5ex

\subsection A.1 Dirac operator and renormalization constants

The lattice theory considered in this paper is 
set up as usual, using the Wilson gauge
action and the standard O($a$)-improved Wilson--Dirac operator.
The notation and normalizations are as in ref.~[\ref{OaImp}].
In particular, $\csw$ and $\cA$ denote the coefficients of
the Pauli term in the Dirac operator
and the O($a$) term required for the improvement of the 
axial current.
Here these coefficients were set to the values
given by the non-perturbatively determined interpolation
formula quoted
in ref.~[\ref{NonPertImp}] (see table~3).

\input table3

The values of the renormalization constant $\ZA$ of 
the axial current listed in table~3
were obtained by evaluating the interpolation formula  
given in ref.~[\ref{NonPertRenI}].
In the case of the renormalization constant $\ZP$ of the 
pseudo-scalar quark density,
the quoted values are the ones
required to pass from the lattice normalization 
of the density to the one in
the $\MSbar$ scheme of dimensional regularization at 
normalization scale $\mu=2$ GeV. 
The constant was calculated
in two steps, first passing from the lattice to the 
renormalization-group-invariant normalization [\ref{NonPertRenII}]
and then from there 
to the $\MSbar$ scheme [\ref{NonPertRenIII}].

\subsection A.2 Quark masses

The renormalized quark mass in the $\MSbar$ scheme is given by
\equation{
  \mr={\ZA(1+\bA a\mq)\over\ZP(1+\bP a\mq)}m+\rmO(a^2),
  \enum
}
where $m$ is the bare current-quark mass, $\mq=m_0-\mc$ the 
subtracted bare mass and $\mc$ the critical bare mass.
Here and below, $b_X$ (where $X=A,P,\ldots$)
denotes an improvement coefficient required
to cancel lattice effects proportional to $a\mq$.

At fixed gauge coupling, the current quark mass is
related to the subtracted bare mass through
\equation{
  m=Z\mq\left\{1+(\bm-\bA+\bP)a\mq\right\}+\rmO(a^2).
  \enum
}
Using the Schr\"odinger functional, the coefficients 
$Z$ and $\bm-\bA+\bP$
were determined non-perturbatively by 
Guagnelli et al.~[\ref{CurrentBareMass}] (see table~4).
Also shown in table~4 is the current quark mass 
at one value of the hopping parameter $\kappa=(8+2am_0)^{-1}$.
Together with eq.~(A.2), these data allow
the current quark mass to be estimated at larger values of 
$\kappa$,
where a direct computation on the lattices considered in this
paper tends to be compromised by the presence
of accidental near-zero modes of the Dirac operator.

\input table4

\subsection A.3 Renormalization of the mode number

The renormalization and improvement properties of the mode number
\equation{
  \nu(M,\mq)=\Tr\{\PM\}
  \enum
}
were discussed in detail in ref.~[\ref{TopProj}]. In
particular, it was shown there that
\equation{
  \nur(\Mr,\mr)=\nu(M,\mq)
  \enum
}
is a renormalized and O($a$)-improved quantity. 
In this equation, the bare parameters $M,\mq$ are to be 
expressed through the renormalized masses 
\equation{
  \Mr=\ZP^{-1}(1+\bmu a\mq)M
  \enum
}
and $\mr$.
Note that $\Mr$ does not renormalize in the same way as the quark mass.
Currently the coefficient 
\equation{
  \bmu=-\frac{1}{2}-0.111(4)\times g_0^2+\rmO(g_0^4),
  \enum
}
is only known to one-loop order of perturbation theory
[\ref{tmQCDimp},\ref{TopProj}].

%% file: table3
\topinsert
\newdimen\digitwidth
\setbox0=\hbox{\rm 0}
\digitwidth=\wd0
\catcode`@=\active
\def@{\kern\digitwidth}
\tablecaption{Improvement coefficients and renormalization constants} 
\vskip1.0ex
$$\vbox{\settabs\+&%
                  xxxxxx&xx&
                  xxxxxxxxx&xx&
                  xxxxxxxxx&xx&
                  xxxxxxxxx&xx&
                  xxxxxxxxx&x\cr
\thicktablerule
\vskip1.0ex
                \+& \hfill $\beta$\hfill
                 && \hfill $\csw$\hfill
                 && \hfill $\cA$\hfill
                 && \hfill $\ZA$\hfill
                 && \hfill $\ZP$\hfill
                 &\cr
\vskip1.0ex
\thintablerule
\vskip1.2ex
  \+& \hfill $5.96$\hfill
  &&  \hfill $1.81663$\hfill
  &&  \hfill $-0.11432$ \hfill
  &&  \hfill $0.789(8)$ \hfill
  &&  \hfill $0.629(14)$ \hfill
  &\cr
\vskip0.3ex
  \+& \hfill $6.17$\hfill
  &&  \hfill $1.63125$\hfill
  &&  \hfill $-0.04015$ \hfill
  &&  \hfill $0.807(8)$ \hfill
  &&  \hfill $0.622(14)$ \hfill
  &\cr
\vskip0.3ex
  \+& \hfill $6.42$\hfill
  &&  \hfill $1.51877$\hfill
  &&  \hfill $-0.02446$ \hfill
  &&  \hfill $0.824(8)$ \hfill
  &&  \hfill $0.618(13)$ \hfill
  &\cr
\vskip1.2ex
\thicktablerule
}
$$
\endinsert

%% file: table4
\topinsert
\newdimen\digitwidth
\setbox0=\hbox{\rm 0}
\digitwidth=\wd0
\catcode`@=\active
\def@{\kern\digitwidth}
\tablecaption{Quark mass reference point and extrapolation coefficients} 
\vskip1.0ex
$$\vbox{\settabs\+&%
                  xxxxxx&xx&
                  xxxxxxxxx&xx&
                  xxxxxxxxxxx&xx&
                  xxxxxxxxxx&x&
                  xxxxxxxxxxxx&x\cr
\thicktablerule
\vskip1.0ex
                \+& \hfill $\beta$\hfill
                 && \hfill $\kappa$\hfill
                 && \hfill $am$\hfill
                 && \hfill $Z$\hfill
                 && \hfill $\bm-\bA+\bP$\hfill
                 &\cr
\vskip1.0ex
\thintablerule
\vskip1.2ex
  \+& \hfill $5.96$ \hfill
  &&  \hfill $0.13360$ \hfill
  &&  \hfill $0.033216(60)$ \hfill
  &&  \hfill $1.0402(4)$ \hfill
  &&  \hfill $-1.017(13)$ \hfill
  &\cr
\vskip0.3ex
  \+& \hfill $6.17$ \hfill
  &&  \hfill $0.13521$ \hfill
  &&  \hfill $0.016636(39)$ \hfill
  &&  \hfill $1.0935(4)$ \hfill
  &&  \hfill $-0.739(13)$ \hfill
  &\cr
\vskip0.3ex
  \+& \hfill $6.42$ \hfill
  &&  \hfill $0.13545$ \hfill
  &&  \hfill $0.008046(20)$ \hfill
  &&  \hfill $1.1041(4)$ \hfill
  &&  \hfill $-0.687(13)$ \hfill
  &\cr
\vskip1.2ex
\thicktablerule
}
$$
\endinsert

%% file: appb
\appendix B. Definition of $\RM$ 

The operator $\RM$ is of the form [\ref{TopProj}]
\equation{
  \RM=h(\Xop),
  \qquad
  \Xop=
  1-{2\Mstar^2\over\Dmdag\Dm+\Mstar^2},
  \enum
}
where $\Mstar\simeq M$ and $h(x)$ is a polynomial approximation
to the step function $\theta(-x)$ in the range $-1\leq x\leq 1$.
More precisely, the polynomial is given by
\equation{
  h(x)=\frac{1}{2}\left\{1-xP(x^2)\right\},
  \enum
}
$P(y)$ being
the polynomial of degree $n$ which
minimizes the deviation
\equation{
  \delta=\max_{\eps\leq y\leq 1}\left|1-\sqrt{y}P(y)\right|
  \enum
}
for some specified (small) value of $\eps$. 
This choice ensures
that $h(x)$ provides a uniform approximation to the step function
in the range $|x|\geq\sqrt{\eps}$, with maximal absolute
deviation equal to $\frac{1}{2}\delta$. 
Moreover, inspection shows
that $h(x)$ decreases monotonically in the
transition region $-\sqrt{\eps}\leq x\leq\sqrt{\eps}$.

For a given degree $n$ and transition range
$\eps$, the coefficients of the minmax polynomial $P(y)$ 
can be computed numerically using standard techniques.
An efficient procedure was described in ref.~[\ref{NumMethods}],
for example. 
The mass $\Mstar\propto M$ is then determined through
\equation{
  {M\over\Mstar}=
  \left({1-\sqrt{\eps}\over1+\sqrt{\eps}}\right)^{1/2}
  +\int_{-\sqrt{\eps}}^{\sqrt{\eps}}
  \rmd x\,{1+x\over(1-x^2)^{3/2}}\,h(x)^4=1+\rmO(\sqrt{\eps}).
  \enum
}
As explained in appendix B of ref.~[\ref{TopProj}],
this convention is intended to minimize the deviation 
$\left|\Tr\{\PM-\RM^4\}\right|$.
In the present context,
other choices of $\Mstar$ would however do just as well,
since eq.~(2.2) is expected to hold for any $M$.

Small approximation errors $\delta$ are achieved
with moderately high degrees $n$ if $\eps$ is not
very small. 
For $n=32$ and $\eps=0.01$, for example, one obtains
\equation{
  \delta=4.37\times10^{-4},
  \qquad
  M/\Mstar=0.96334.
  \enum
}
The transition range $|x|\leq\sqrt{\eps}$ approximately
corresponds to the range 
\equation{
  0.9\leq\sqrt{\alpha}/\Mstar\leq1.1
  \enum
} 
of eigenvalues $\alpha$ of $\Dmdag\Dm$ in this case.

%% file: biblio
\beginbibliography


\bibitem{ChiWI}
L. Giusti, G. C. Rossi, M. Testa, G. Veneziano,
{\it The $U_A(1)$ problem on the lattice with Ginsparg--Wilson fermions},
Nucl. Phys. B628 (2002) 234

\bibitem{ChiWII}
L. Giusti, G. C. Rossi, M. Testa,
{\it Topological susceptibility in full QCD with Ginsparg--Wilson fermions},
Phys. Lett. B587 (2004) 157


\bibitem{TopSing}
M. L\"uscher,
{\it Topological effects in QCD and the problem of short-distance 
singularities},
Phys. Lett. B593 (2004) 296


\bibitem{TopProj}
L. Giusti, M. L\"uscher,
{\it Chiral symmetry breaking and the Banks--Casher relation in lattice QCD 
with Wilson quarks},
JHEP 03 (2009) 013


\bibitem{Wilson}
K. G. Wilson, {\it Confinement of quarks},
Phys. Rev. D10 (1974) 2445


\bibitem{SW}
B. Sheikholeslami, R. Wohlert,
{\it Improved continuum limit lattice action for QCD with Wilson fermions},
Nucl. Phys. B259 (1985) 572

\bibitem{OaImp}
M. L\"uscher, S. Sint, R. Sommer, P. Weisz,
{\it Chiral symmetry and O(a) improvement in lattice QCD},
Nucl. Phys. B478 (1996) 365


\bibitem{OsbornEtAl}
J. C. Osborn, D. Toublan, J. J. M. Verbaarschot,
{\it From chiral random matrix theory to chiral perturbation theory},
Nucl. Phys. B540 (1999) 317


\bibitem{DelDebbioTauQ}
L. Del Debbio, H. Panagopoulos, E. Vicari,
{\it $\theta$-dependence of SU(N) gauge theories},
JHEP 08 (2002) 044


\bibitem{StefanConference}
S. Schaefer, R. Sommer, F. Virotta,
{\it Investigating the critical slowing down of QCD simulations},
PoS (LAT2009) 032


\bibitem{SommerScale}
R. Sommer,
{\it A new way to set the energy scale in lattice gauge theories 
and its applications to the static force and $\alpha_s$ in 
SU(2) Yang--Mills theory},
Nucl. Phys. B411 (1994) 839

\bibitem{GuagnelliEtAl}
M. Guagnelli, R. Sommer, H. Wittig (ALPHA collab.),
{\it Precision computation of a low-energy reference scale in 
quenched lattice QCD},
Nucl. Phys. B535 (1998) 389


\bibitem{MichaelPeisa}
C. Michael, J. Peisa,
{\it Maximal variance reduction for stochastic propagators with 
applications to the static quark spectrum},
Phys. Rev. D58 (1998) 034506


\bibitem{LocDefl}
M. L\"uscher,
{\it Local coherence and deflation of the low quark modes in lattice QCD},
JHEP 07 (2007) 081


\bibitem{CompStrat}
M. L\"uscher,
{\it Computational Strategies in Lattice QCD}, 
Lectures given at the Summer School on ``Modern perspectives in lattice QCD'', 
Les Houches, August 3-28, 2009, 
arXiv:1002.4232v2 [hep-lat]


\bibitem{ChiGW}
L. Del Debbio, L. Giusti, C. Pica,
{\it Topological susceptibility in SU(3) gauge theory},
Phys. Rev. Lett. 94 (2005) 032003

\bibitem{IndThm}
P. Hasenfratz, V. Laliena, F. Niedermayer,
{\it The index theorem in QCD with a finite cutoff},
Phys. Lett. B427 (1998) 125


\bibitem{WilsonFlow}
M. L\"uscher,
{\it Properties and uses of the Wilson flow in lattice QCD},
JHEP 08 (2010) 071


\bibitem{NonPertImp}
M. L\"uscher, S. Sint, R. Sommer, P. Weisz, U. Wolff,
{\it Nonperturbative O(a) improvement of lattice QCD},
Nucl. Phys. B491 (1997) 323


\bibitem{NonPertRenI}
M. L\"uscher, S. Sint, R. Sommer, H. Wittig
(ALPHA Collab.),
{\it Nonperturbative determination of the axial current 
normalization constant in O(a) improved lattice QCD},
Nucl. Phys. B491 (1997) 334

\bibitem{NonPertRenII}
S. Capitani, M. L\"uscher, R. Sommer, H. Wittig
(ALPHA Collab.),
{\it Nonperturbative quark mass renormalization in quenched lattice QCD},
Nucl. Phys. B544 (1999) 669

\bibitem{NonPertRenIII}
J. Garden, J. Heitger, R. Sommer, H. Wittig 
(ALPHA and UKQCD Collab.),
{\it Precision computation of the strange quark's mass in quenched QCD},
Nucl. Phys. B571 (2000) 237


\bibitem{CurrentBareMass}
M. Guagnelli, R. Petronzio, J. Rolf, S. Sint, R. Sommer, U. Wolff
(ALPHA Collab.),
{\it Non-perturbative results for the coefficients 
$\bm$ and $\bA-\bP$ in O(a) improved lattice QCD},
Nucl. Phys. B595 (2001) 44


\bibitem{tmQCDimp}
R. Frezzotti, S. Sint, P. Weisz (ALPHA Collab.),
{\it O(a) improved twisted mass lattice QCD},
JHEP 07 (2001) 048


\bibitem{NumMethods}
L. Giusti, C. Hoelbling, M. L\"uscher, H. Wittig,
{\it Numerical techniques for lattice QCD in the epsilon regime},
Comput. Phys. Commun. 153 (2003) 31

\endbibliography